\documentstyle[11pt,epsf]{article}
\begin{document}

{\small
\rightline{\it Contribution to the} 
\vspace{0.5pc}
\rightline{\it 4th International Seminar 
on Interaction of Neutrons with Nuclei}
\rightline{\it ``Neutron Spectroscopy, Nuclear Structure, Related Topics''}
\rightline{\it Dubna (Russia), April 1996} 
\vspace{1.5pc}}

\centerline{\Large\bf EXOTIC PROPERTIES OF LIGHT NUCLEI}
\vspace{1.0pc}
\centerline{\Large\bf AND THEIR NEUTRON CAPTURE CROSS SECTIONS}
\vspace{1.5pc}

\centerline{A. Mengoni${}^{1,2}$, T. Otsuka${}^{1,3}$, 
T. Nakamura${}^{3}$, and M. Ishihara${}^{1,3}$} 
\centerline{\it ${}^1$RIKEN, Radiation Laboratory,
2-1 Hirosawa, Wako, Saitama 351-02, Japan}
\centerline{\it ${}^2$ENEA, Applied Physics Section,
v. Don Fiammelli 2, 40128 Bologna, Italy}
\centerline{\it ${}^3$The University of Tokyo,
Department of Physics, Hongo, Bunkyo-ku, 
Tokyo 113, Japan}

\vspace{1.5pc}

\begin{abstract}
We have investigated the implications 
of the neutron halo configuration, observed
in the ground-state of some neutron-rich
light nuclei, on neutron radiative transition
processes. In particular, we have studied the
influence of the neutron halo 
on the direct radiative capture (DRC) process.
The energy dependence as well as
the strength of E1 emission due to
incident {\it p}-wave neutrons is
strongly influenced by the halo configuration 
of the residual nucleus capturing state.
We have compared the calculated
${}^{10}$Be$(n,\gamma){}^{11}$Be DRC cross section
with that derived from the experiment in the inverse 
kinematics (Coulomb dissociation of ${}^{11}$Be).
We show from the comparison that some
important information on the structure of 
the halo nucleus ${}^{11}$Be can be derived.
\end{abstract}

\section{Introduction}

In the category of ``exotic'' nuclear structure
properties we can include the neutron skin 
and the neutron halo observed in ground-state 
configurations of light neutron rich nuclei \cite{Ta91,Tax92}.
A rigorous definition of neutron skin and of neutron
halo cannot be given. Therefore, we will show here an example 
of both structures to elucidate their features and to show 
their basic properties.
The effect of neutron skin structure 
has been recently investigated \cite{Fux93,Yox95} and 
has been shown to be responsible for a low energy 
excitation mode, decoupled from the isoscalar giant 
quadrupole resonance \cite{Yox95}.
Here we will concentrate on the effects of the neutron 
halo structure. In particular, we will show its importance 
in relation to the low energy electric dipole transition 
mechanism in ${}^{11}$Be.

\subsection{Neutron skin}
\begin{figure}[t]
\begin{center}
\leavevmode
\hbox{
\epsfxsize=10.0cm
\epsfysize=9.0cm
\epsffile{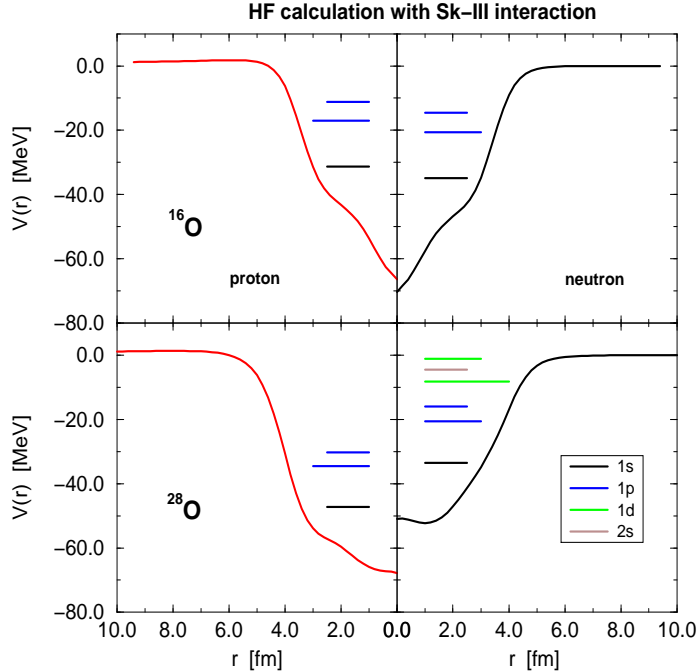}}
\end{center}
\caption{Mean-field calculation for ${}^{16}$O (upper) and
${}^{28}$O (lower). Proton and neutron potentials are
shown respectively on the left and right side.
}
\label{fig:O_potential}
\end{figure}

An example of neutron skin structure is shown 
in Figure \ref{fig:O_potential}.
There, the result of a mean-field calculation 
using Hartree-Fock method \cite{Re91} is shown 
for the two nuclei ${}^{16}$O (in the upper part) 
and ${}^{28}$O (in the lower part). 
The calculations were performed using the 
common Skyrme-S3 interaction, but the general features 
of the result can be deduced with any other kind 
of effective interaction. 
In the upper part of the figure,
the mean-field potential for protons (left) 
and neutrons (right) is shown together with the bound 
single-particle energy spectrum. In this double-magic
nucleus, neutrons and protons occupy analogous 
single-particle orbits located at comparable energies.

%
%
%
%
%
%
\begin{figure}[t]
\begin{center}
\leavevmode
\hbox{
\epsfxsize=10.0cm
\epsfysize=9.0cm
\epsffile{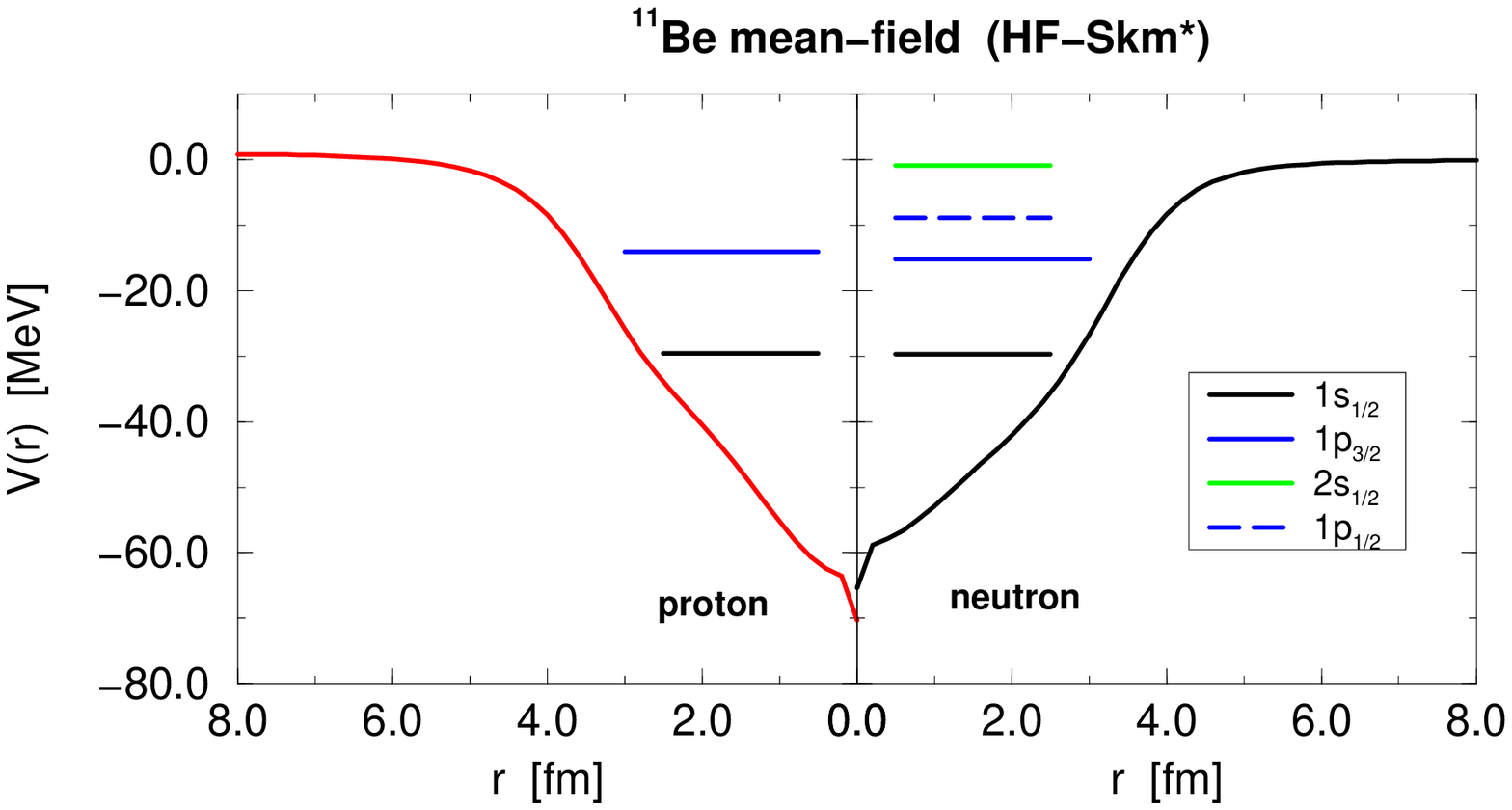}
}
\end{center}
\vspace*{-10pt}
\caption{Mean-field calculation for ${}^{11}$Be.
Proton and neutron potentials are
shown respectively on the left and right side.
Note that the $2s_{1/2}$ state is locate at the
top of the potential well (it is bound by only
505 keV). In a more realistic description, 
the $1p_{1/2}$ orbit, here forced to be 
unoccupied, should be 320 keV above the 
ground-state (See Figure 4 below).}
\label{fig:11Be_potential}
\end{figure}
%
%
In the lower part of the figure, the same type
of calculation is shown for the very neutron rich 
${}^{28}$O (N=20) nucleus, a double-magic nucleus too. 
In this case the neutrons occupy single-particle states 
with energies very different compared to the proton 
case.
Moreover, the Fermi energies of protons and neutrons 
are very different here due to the much larger neutron
number.
A plot of the densities for these two nuclei would reveal 
a thick neutron configuration ``covering'' the proton 
density distribution.

A typical shell model configuration can be seen in the 
mean-field calculation shown in Figure \ref{fig:O_potential}. 
In the case of ${}^{28}$O, the {\it s} and {\it d} shells are 
both occupied and the single-particle ordering of 
the states is the usual $1d_{5/2}, 2s_{1/2}, 1d_{3/2}$. 
The last occupied orbit is therefore the $1d_{3/2}$ orbit. 

\subsection{Neutron halo}
A different situation is encountered when similar calculations
are performed for the neutron rich ${}^{11}$Be nucleus. 
It is well known \cite{Aj90} that its ground state 
is a $J^{\pi}=1/2^{+}$ state with a dominant 
\mbox{$ | {}^{10}\mbox{Be}(0^{+}) \otimes (2s_{1/2})_{\nu} >$}
configuration.
This is in contrast with the normal single-particle ordering, 
where the $1p_{1/2}$ orbit is occupied before the $2s_{1/2}$ 
orbit. This peculiar configuration, together with the very low 
binding energy, 0.505 MeV, builds up the neutron halo structure 
of the ${}^{11}$Be ground state. 

\begin{figure}[t]
\begin{center}
\leavevmode
\hbox{
\epsfxsize=10.0cm
\epsfysize=8.0cm
\epsffile{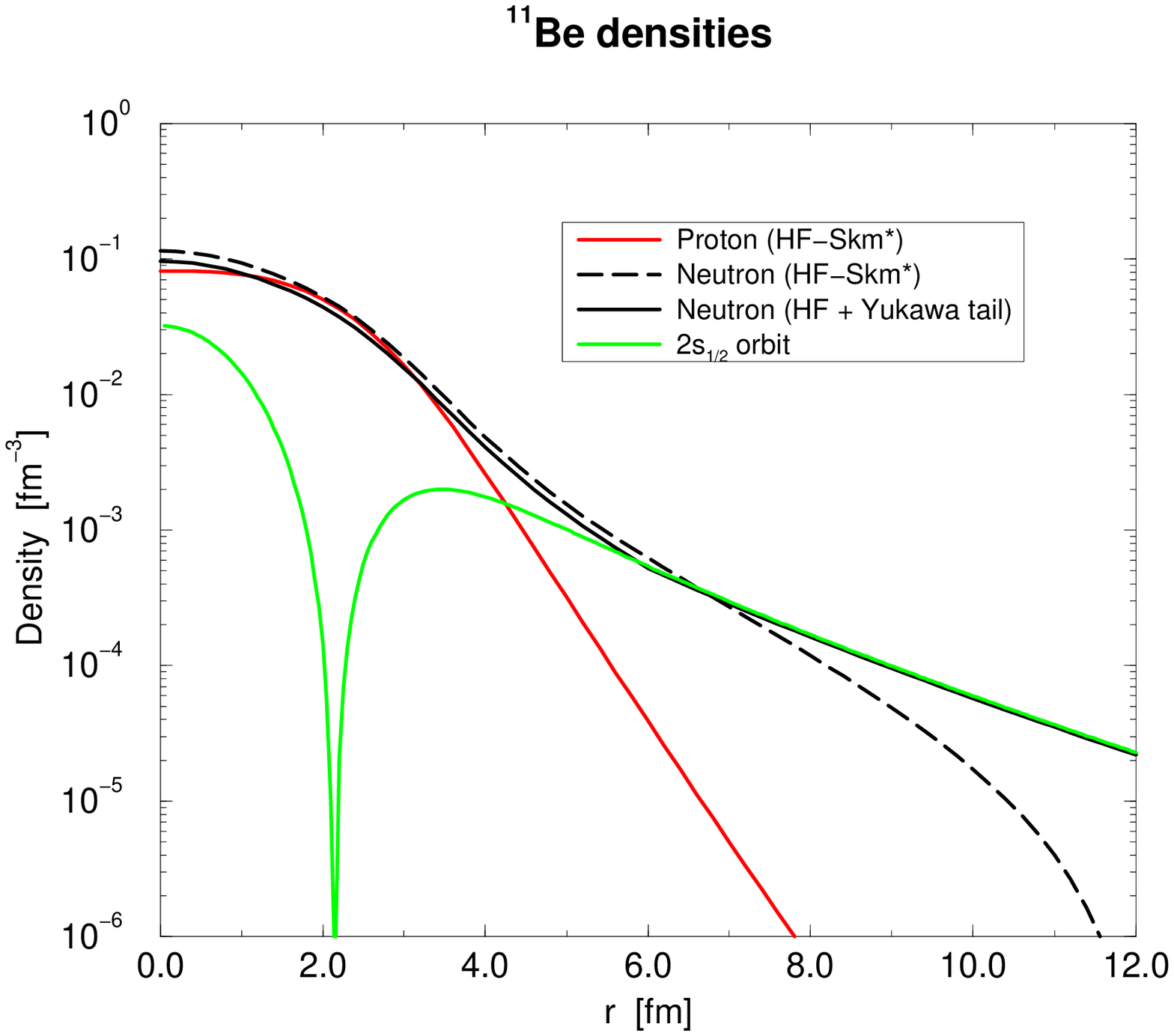}
}
\end{center}
\vspace*{-10pt}
\caption{Mean-field calculation for ${}^{11}$Be.
The neutron density corresponding to the $2s_{1/2}$
single-particle state bound by 0.505 MeV is shown
for comparison with the Hartree-Fock results. 
}
\label{fig:11Be_density}
\end{figure}

The mean-field potential for ${}^{11}$Be is shown 
in Figure \ref{fig:11Be_potential}.
The tail component of the corresponding neutron density, 
extending well outside the nuclear surface (defined by a 
root-mean-square radius of 
$2.86 \pm 0.04$ fm \cite{Nax94}) can be seen 
in Figure \ref{fig:11Be_density}. 
%
%
%
%
There, the neutron and proton densities are shown as calculated by the
Hartree-Fock method with a Skyrme-m* interaction. The standard 
Hartree-Fock calculation would produce a ground-state with a
$1p_{1/2}$ configuration. To simulate the experimental
situation, the results shown in the figure
are obtained forcing the ground-state to be 
a $2s_{1/2}$ state. Also shown in the figure is a neutron density 
with an added Yukawa-tail determined by fixing the binding energy of 
the ground-state. In this way, the resulting 
neutron density compares well with the experimentally derived
density. It should be noted that the wave function corresponding 
to the $2s_{1/2}$ orbit in a Woods-Saxon potential itself 
can reproduce the external part of the neutron density.
%
%
\begin{figure}[t]
\begin{center}
\leavevmode
\hbox{
\epsfxsize=9.0cm
\epsfysize=5.0cm
\epsffile{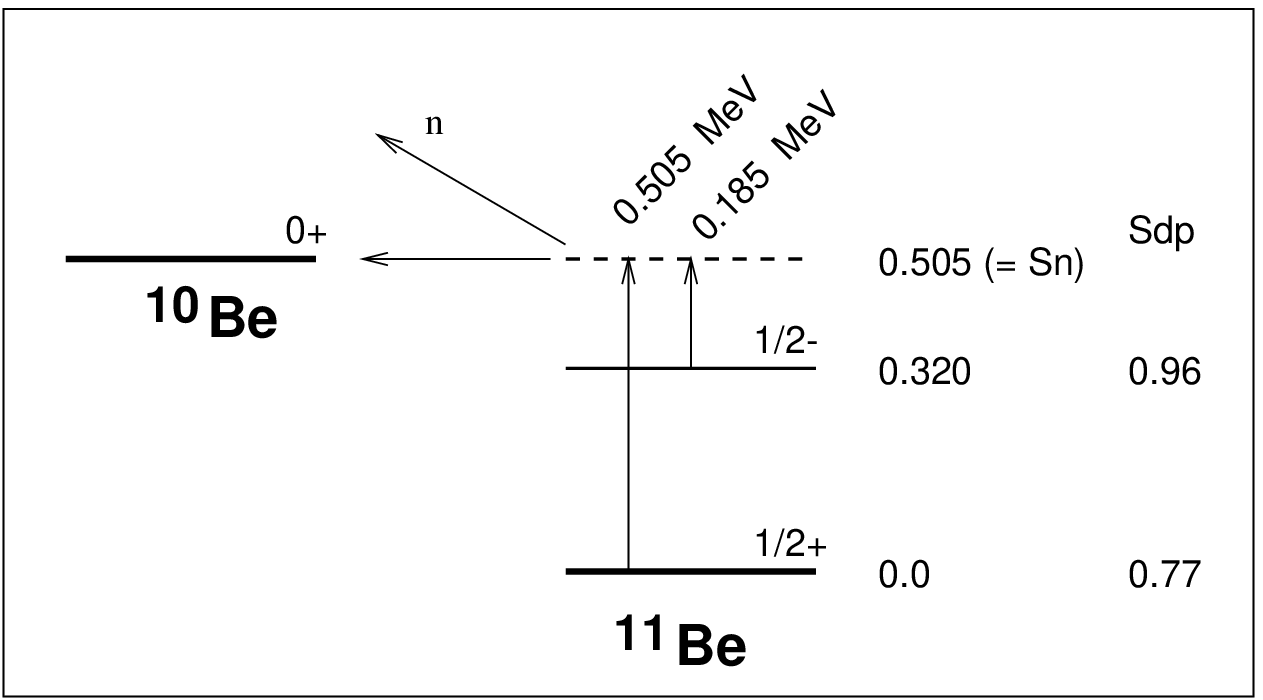}
}
\end{center}
\vspace*{-10pt}
\caption{The Coulomb dissociation of ${}^{11}$Be
into ${}^{10}$Be$ + n$. The spectroscopic factors
of the two bound states are those derived from
a ${}^{10}$Be$(d,p)$ measurement \protect\cite{Aj90}.
}
\label{fig:11Be_diss}
\end{figure}
%
%
In the calculations of the capture cross section shown
below we have used this wave function to evaluate the
electric dipole matrix elements.

\section{Neutron direct radiative transitions}
The peculiar structure properties briefly described
in the previous section have been shown to exhibit
strong influence on nuclear excitation modes 
\cite{Fux93,Yox95,Otx94,Mex95}. 
In what follows, we will show the results of the effect 
of the neutron halo structure on electric dipole transitions.
These can be investigated, in principle, 
in both the reaction directions: in the neutron 
capture as well as in the Coulomb dissociation process.

Our group has recently investigated the effect of
the nuclear wave function component in the
external region on the neutron capture
process \cite{Otx94,Mex95}. 
We have calculated several neutron capture
cross sections of light nuclei using the direct radiative
capture (DRC) model. 
The calculations for $^{12}$C$(n,\gamma)$ and
$^{16}$O$(n,\gamma)$ reactions have been compared
with recent experimental results from direct
measurements \cite{Mex95,Padova94,RIKEN-INFN95}.

The DRC reaction mechanism may be responsible
for the most part of the capture reaction
cross section in the particular condition in which
the density of states is low enough to
hinder the compound nucleus formation.
Furthermore, because
the halo structure arises mainly
from loosely bound {\it s} orbits (see the discussion
above), electric dipole $\gamma$-ray
emissions can only be induced by
incident {\it p}-wave neutrons.
In fact, the DRC process is essentially determined 
by the E1 transition matrix elements
$$
{{\it Q^{(E1)}_{i \rightarrow f}}} = < \Psi _{f}
\vert \hat{T}^{E1} \vert \Psi _{i} >.
$$ 
In the case of incident {\it p}-wave neutrons, 
these matrix elements are very much sensitive 
to the tail component of the final capturing state 
wave function $ \Psi _{f}$ and very little sensitive 
to the treatment of the incident channel neutron 
scattering state $\Psi _{i}$ \cite{Mex95}.
The energy dependence as well as
the strength of E1 emission due to
incident {\it p}-wave neutrons is
therefore strongly influenced by
the neutron halo structure of
the residual nucleus capturing state.
Whether this state is the ground
or an excited nuclear state makes
no difference in this scheme.
%
%
\begin{figure}[t]
\begin{center}
\leavevmode
\hbox{
\epsfxsize=10.0cm
\epsfysize=9.0cm
\epsffile{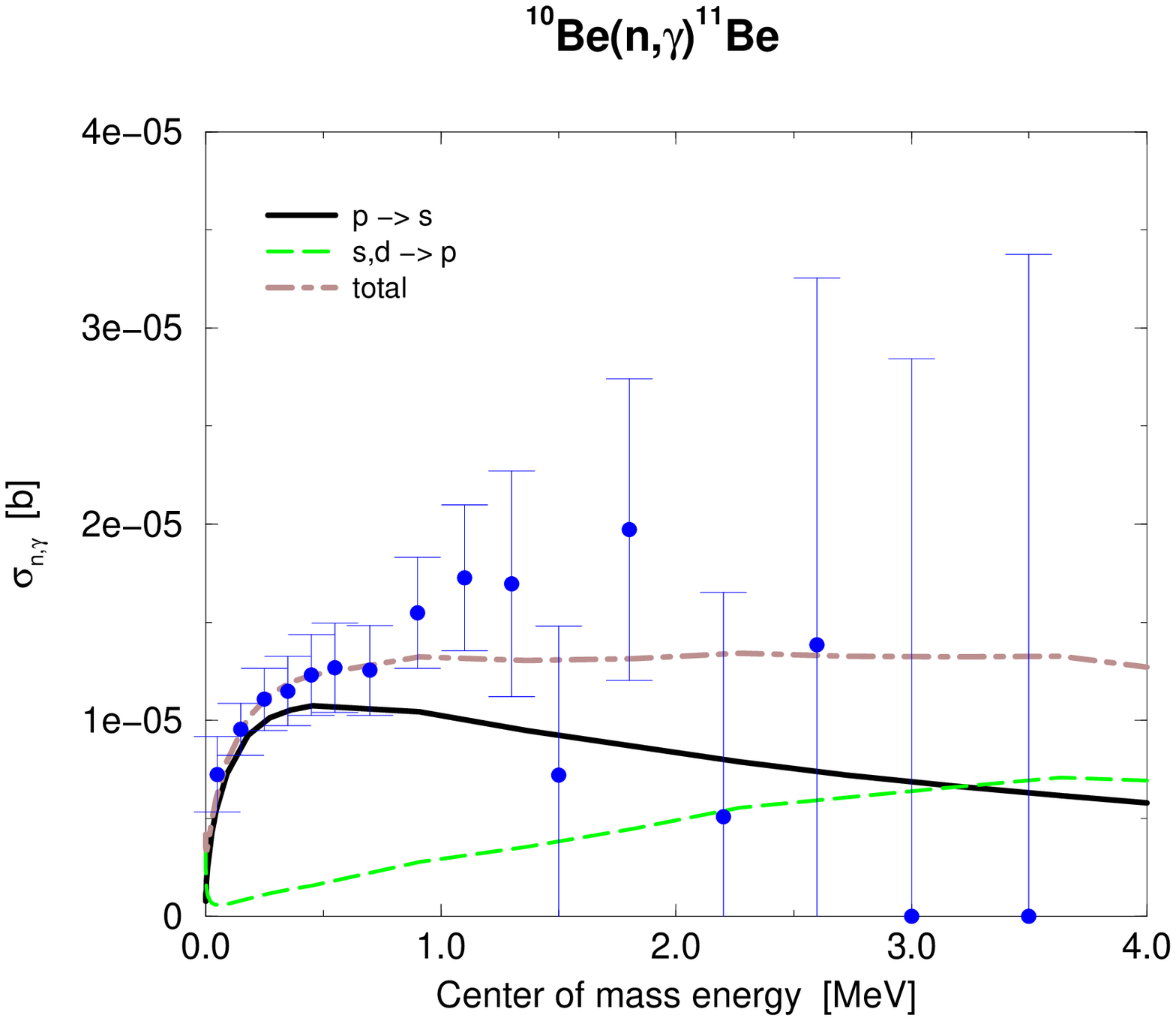}
}
\end{center}
\vspace*{-10pt}
\caption{Neutron capture cross section
of ${}^{10}$Be as a function of the center of mass
energy. The experimental values are deduced, as
described in the text, from a Coulomb dissociation 
experiment \protect\cite{Nax94}. The large uncertainty
on the experimental values at high energy is amplified 
by the rapid decrease of the B(E1) strength distribution 
with increasing excitation energy.
}
\label{fig:10Be_ng}
\end{figure}
%
%
As mentioned above, we can compare the
${}^{10}$Be$(n,\gamma){}^{11}$Be DRC cross section
with that derived from the experiment in the inverse
kinematics: the Coulomb dissociation of ${}^{11}$Be.
A recent Coulomb dissociation experiment has been performed 
using the radioactive beam line RIPS at RIKEN \cite{Nax94}.
A schematic view of the process is shown 
in Figure \ref{fig:11Be_diss}.
Note that in the actual experiment, the dissociation from
the first excited state at $E_{x} = 0.320$ MeV 
($J^{\pi} = 1/2^{-}$) does not take place because
all the nuclei of the incident beam are in their
ground state.

The electric dipole strength distribution $dB(E1)/dE_x$,
as measured in the Coulomb dissociation experiment, 
can be related to the neutron capture cross section,
$\sigma_{n,\gamma}^{E1}$, by
$$
\frac{dB(E1)}{dE_x} = \frac{9}{8 \pi^3} \ 
\Bigl( \frac{\hbar c}{E_x} \Bigr)^3 \ 
{k_{n}^2} \ \sigma_{n,\gamma}^{E1}(E_x) 
$$
where $E_x$ is the excitation energy and $k_n$ the
emitted neutron wave number. 
This relation is only valid for
the present combination of angular momentum values.

The results of the DRC calculation in comparison to
the experimental values derived from the Coulomb
dissociation experiment and converted into the
neutron capture channel are shown in Figure \ref{fig:10Be_ng}. 
As a first comment to this result, one has to note
that the experimental values have to be compared
only to the calculation including the $p \rightarrow s$
transition because, as noted above, the experiment
only includes incident ${}^{11}$Be nuclei in their
ground-state.

The next feature which can be seen in the Figure is that
the low energy part, where the experimental uncertainty
is lowest, is quite well reproduced by the
calculation. This can be interpreted as a confirmation
of the halo structure of the ground-state of ${}^{11}$Be.
In fact, the matrix elements which enter in the expression
for the DRC capture given above can be decomposed into
a product of three factors
$$
Q^{(E1)}_{i \rightarrow f}
\equiv \sqrt{S_f} \ {\cal I}_{i,f} \ A_{i,f}
$$
where $S_f$ is the spectroscopic factor of the bound
state, $A_{i,f}$ a geometrical factor containing only
angular momentum coupling constants and ${\cal I}_{i,f}$
is the radial part of the overlap between the initial and
final state. Now, if the initial state only consists
of {\it p}-wave neutrons in the continuum, its wave 
function is simply given by 
$$
\Psi _{l m}(\mbox{\bf r}) \equiv
w_{l}(r)
\frac{Y_{l,m}(\theta,\phi)}{r v^{1/2}}
$$
with $w_{l=1}(r) \propto j_{1}(k_{n}r)$ ($j_{1}$ is a
spherical Bessel function). The DRC cross section is
proportional to the square of the overlap integral 
$$
{\cal I}_{l_{i}l_{f}} \equiv \int\limits_{0}^{\infty}
u_{l_f}(r) r w_{l_i}(r) dr
$$
and is, therefore, nothing but a Fourier-Bessel transformation
of the bound state radial wave function $u_{l=0}(r)$.
A measure of the capture cross section (or of the
Coulomb dissociation cross section) is a measure of the
wave function in the Fourier space. Then,
the low energy  part of the capture cross section plot
corresponds to large values of the radial coordinate 
which is influenced by the halo configuration.

Finally, the fluctuations in the measured values at 
$E_{x} \approx 2$ MeV, though still masked by a large 
experimental uncertainty, may contain information 
on a ${}^{11}$Be excited state. 
In fact, the first state above the neutron emission
threshold is at $E_{x} = 1.778 $ MeV (corresponding
to $E_{cm} = 1.273 $ MeV) \cite{Aj90} and has 
$J^{\pi}=3/2^{+}$ or $J^{\pi}=5/2^{+}$. It can be populated 
by electric quadrupole excitation with the emission 
of {\it d}-wave neutrons. Should the experimental
uncertainty be reduced to the level of the low energy
part, the structure of this ${}^{11}$Be excited state
could be deduced.

\section{Conclusion}

We have shown that the halo structure of light neutron-rich
nuclei can be investigated in the frame of a simple
DRC model. We have shown this in the particular case
in which the cross section was not obtained in a
direct measurement but derived from an experiment
in the inverse kinematics. This technique can be
very useful to study ``exotic'' nuclear
structure properties. In addition, it can be
utilized to obtain neutron capture cross sections
for unstable nuclei where direct measurements are
not possible. This possibility is particularly
important for applications in nuclear
astrophysics \cite{Atami96} where neutron capture reaction
rates are needed to investigate primordial
(big-bang) as well as stellar nucleosynthesis.

\section*{Acknowledgments}

We thank Y. Nagai, T. Shima and N. Fukunishi
for many useful discussions that greatly helped to
generate the present contribution. This work has been
supported by the Japan Science and Technology
Agency through the STA fellowship program
(ID 194102).

\end{document}